\documentclass[pdftex,twocolumn,epjc3]{svjour3}          

\RequirePackage[T1]{fontenc}

\smartqed  

\RequirePackage{graphicx}
\RequirePackage{flushend}
\RequirePackage[numbers,sort&compress]{natbib}
\RequirePackage[colorlinks,citecolor=blue,urlcolor=blue,linkcolor=blue]{hyperref}
\usepackage{amsmath}
\usepackage{lineno}
\usepackage{multirow}
\usepackage{makecell}
\usepackage{xspace}
\usepackage{hhline}
\usepackage{amsfonts}
\usepackage{tabularx}
\usepackage{cuted}
\usepackage{amssymb}
\usepackage{amsmath}
\journalname{Eur. Phys. J. C}

\begin{document}

\title{\boldmath Quark/Gluon Discrimination and Top Tagging with Dual Attention Transformer}

\author{Minxuan He\thanksref{e1,addr1,addr2}
        \and
        Daohan Wang\thanksref{e2,addr3} 
}

\thankstext{e1}{e-mail: hemx@amss.ac.cn}
\thankstext{e2}{e-mail: wdh9508@gmail.com}

\institute{University of Chinese Academy of Sciences, Beijing 100049, PR China\label{addr2}
          \and
          Academy of Mathematics and Systems Science, Chinese Academy of Sciences, Beijing 100190, PR China\label{addr1}
          \and
          Department of Physics, Konkuk University, Seoul 05029, Republic of Korea\label{addr3}
}

\date{Received: date / Accepted: date}

\maketitle

\begin{abstract}
Jet tagging is a crucial classification task in high energy physics. Recently the performance of jet tagging has been significantly improved by the application of deep learning techniques. In this study, we introduce a new architecture for jet tagging: the Particle Dual Attention Transformer (P-DAT). This novel transformer architecture stands out by concurrently capturing both global and local information, while maintaining computational efficiency. Regarding the self attention mechanism, we have extended the established attention mechanism between particles to encompass the attention mechanism between particle features. The particle attention module computes particle level interactions across all the particles, while the channel attention module computes attention scores between particle features, which naturally captures jet level interactions by taking all particles into account. These two kinds of attention mechanisms can complement each other. Further, we incorporate both the pairwise particle interactions and the pairwise jet feature interactions in the attention mechanism. We demonstrate the effectiveness of the P-DAT architecture in classic top tagging and quark-gluon discrimination tasks, achieving competitive performance compared to other benchmark strategies.
\end{abstract}

\section{Introduction}
\label{sec:intro}

In high-energy physics experiments, tagging jets, which are collimated sprays of particles produced from high-energy collisions, is a crucial task for discovering new physics beyond the Standard Model. Jet tagging involves distinguishing boosted heavy particle jets from those of QCD initiated quark/gluon jets. Since jets initiated by different particles exhibit different characteristics, two key issues arise: how to represent a jet and how to analyze its representation. Conventionally, jet tagging has been performed using hand-crafted jet substructure variables based on physics motivation. Nevertheless, these methods can often fall short in capturing intricate patterns present in the raw data. 

Over the past decade, deep learning approaches have been extensively adopted to enhance the jet tagging performance\cite{Larkoski:2017jix}. Various jet representations have been proposed, including image-based representation using Convolutional Neural Network (CNN)\cite{Cogan:2014oua,deOliveira:2015xxd,Ren:2021prq,ATLAS:2017dfg,Lin:2018cin,Kasieczka:2017nvn,Macaluso:2018tck,Li:2020grn}, sequence-based representation with Recurrent Neural Network\cite{deLima:2021fwm,ATL-PHYS-PUB-2017-003}, tree-based representation with Recursive Neural Network\cite{Cheng:2017rdo,Louppe:2017ipp} and graph-based representation with Graph Neural Network (GNN)\cite{Ju:2020xty, Ma:2022bvt,Abdughani:2018wrw,Abdughani:2020xfo,Gong:2022lye,Shlomi:2020gdn}. More recently, One representation approach that has gained significant attention is to view the set of constituent particles inside a jet as points in a point cloud. Point clouds are used to represent a set of objects in an unordered manner, described in a defined space. By adopting this approach, each jet can be interpreted as a particle cloud, which treats a jet as a permutation-invariant set of particles, allowing us to extract meaningful information with deep learning method. Based on the particle cloud representation, various deep learning architectures have been introduced, such as Deep Set Framework\cite{Komiske:2018cqr}, ABCNet\cite{Mikuni:2020wpr}, LorentzNet\cite{Gong:2022lye} and ParticleNet\cite{Qu:2019gqs}. Deep Set Framework provides a comprehensive explanation of how to parametrize permutation invariant functions for inputs with variable lengths, taking into consideration both infrared and collinear safety. ParticleNet adapts the Dynamic Graph CNN architecture\cite{wang2019dynamic}, while ABCNet takes advantage of attention mechanisms to enhance the local feature extraction. The LorentzNet focused more on incorporating inductive biases derived from physics principles into the architecture design, utilizing an eﬃcient Minkowski dot product attention mechanism. All of these architectures realize substantial performance improvement on top tagging and quark/gluon discrimination benchmarks.

Over the past few years, attention mechanisms have become as a powerful tool for capturing intricate patterns in sequential and spatial data. The Transformer architecture\cite{vaswani2017attention}, which leverages attention mechanisms, has been highly successful in natural language processing and computer vision tasks such as image recognition. However, when dealing with point cloud representation, which inherently lack a specific order, modifications to the original Transformer structure are required to establish a self-attention operation that is invariant to input permutations. To address these issues, a recent approach known as Point Cloud Transformer (PCT)\cite{Guo_2021,Mikuni:2021pou} was proposed, which entails passing input points through a feature extractor to create a high-dimensional representation of particle features. The transformed data is then passed through a self-attention module that introduces attention coefficients for each pair of particles. Another notable approach is the Particle Transformer\cite{Qu:2022mxj}, which incorporates pairwise particle interactions within the attention mechanism and obtains higher tagging performance than a plain Transformer and surpasses the previous state-of-the-art, ParticleNet, by a large margin.

In recent studies, the Dual Attention Vision Transformer (DaViT)\cite{ding2022davit} has exhibited promising results for image classification. The DaViT introduces the dual attention mechanism, comprising spatial window attention and channel group attention, enabling the effective capture of both global and local features in images. In this paper, we utilize the dual attention mechanism for jet tagging based on point cloud representation. We expanded the particle self-attention established by existing works by introducing the channel self-attention. In the particle self-attention, the particle number defines the scope, and the dimension of particle feature defines the feature dimension. While in the channel self-attention, the channel dimension defines the scope, and the particle number defines the feature dimension. Thus each channel contains an abstract representation of the entire jet. By performing self-attention on these channels, we capture the global interaction by considering all the particles when computing attention scores between each pair of channels. Compared to existing particle self-attention, the channel self-attention is naturally imposed from a global jet perspective rather than a particle one. To achieve the dual attention mechanism, we introduce the Channel Attention module. By alternately applying the Particle Attention module and the Channel Attention module to combine both the local information of the particle representation and the global information of the jet representation for jet tagging, we build a new network structure, called Particle Dual Attention Transformer (P-DAT). Furthermore, inspired by Ref.\cite{Qu:2022mxj}, we design the pairwise jet feature interaction. We incorporate both the pairwise particle interaction and the pairwise jet feature interaction to increase the expressiveness of the attention mechanism. We evaluate the performance of P-DAT on top tagging and quark/gluon discrimination tasks and compare its performance against other baseline models. Our analysis demonstrates the effectiveness of P-DAT in jet tagging and highlights its potential for future applications in high-energy physics experiments.

This article is organized as follows. In Section~\ref{sec:Architecture}, we introduce the Particle Dual Attention Transformer for jet tagging, providing a detailed description of model implementation. In Section~\ref{sec:Classification}, we present the performance of P-DAT and the existing algorithms obtained for top tagging task and quark/gluon discrimination task, utilizing several evaluation metrics and provide an extensive discussion of these results. In Section~\ref{sec:Complexity}, we conduct a comprehensive comparison of computational resource requirements for evaluating each model, including the number of trainable weights and the number of floating-point operations (FLOPs). Finally, our conclusions are presented in Section~\ref{sec:Summary}.

\section{Model Architecture}
\label{sec:Architecture}

The focus of this paper is to introduce the Particle Dual Attention Transformer (P-DAT), which is designed to capture both the local particle-level information and the global jet level information. In this section, we first introduce overall structure of the model architecture. Then we delve into the details of the Channel Attention module and its combination with the Particle Attention module. Finally, we present the model implementation.

\subsection{Overall Structure}

\begin{figure*}[h]
\centering
\includegraphics[width=13cm,height=6cm]{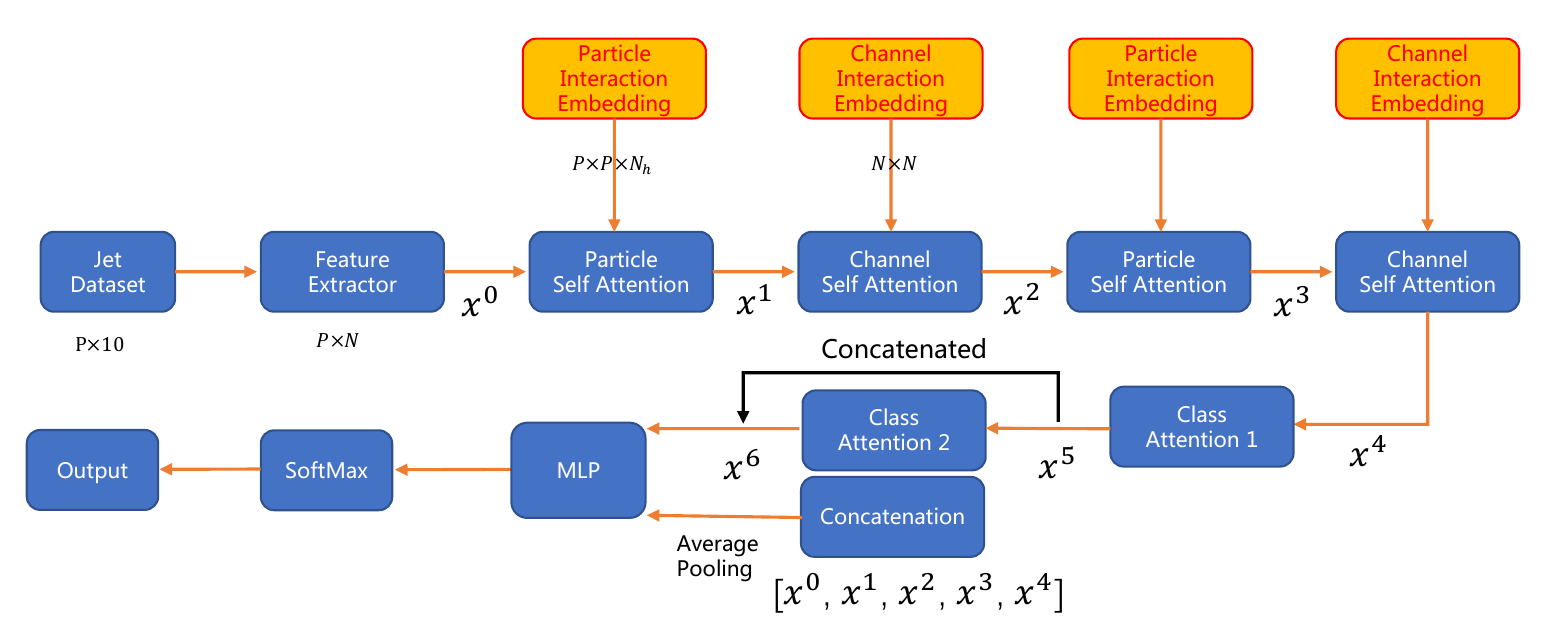}
\vspace{-0.5cm}
\caption{Illustration of the whole model architecture.}
\label{model}
\end{figure*}

The whole model architecture is illustrated in Figure.\ref{model}. It contains three key components, namely the Feature Extractor, the Particle Attention module and the Channel Attention module.

\begin{figure}[ht]
    \centering
    \includegraphics[width=8cm,height=8cm]{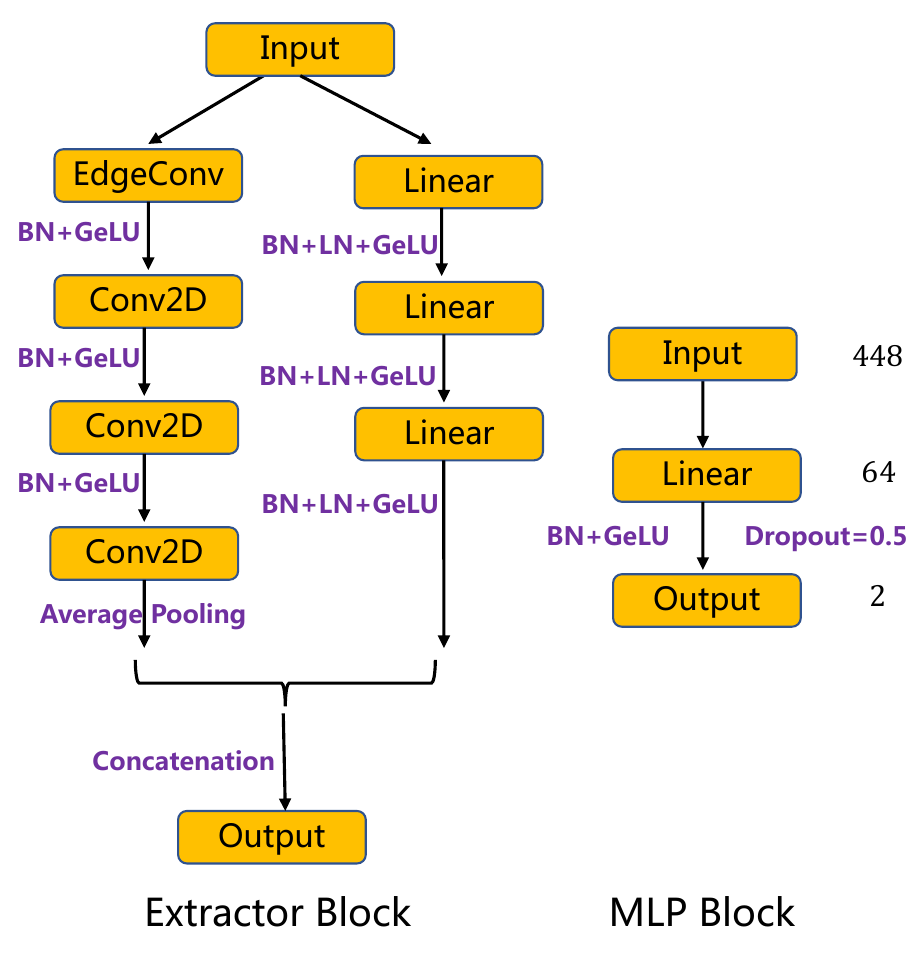}
    \caption{Illustration of the Feature extractor block and the MLP block.}
    \label{part1}
\end{figure}

First of all, we employ the same feature extractor as in Ref.~\cite{Mikuni:2021pou} to transform the inputs from $P\times 7$ to a higher dimensional representation $P\times N$, where P represents the number of particles within the jet, and N denotes the dimension of the embedding features for each particle. As shown in Figure.\ref{part1}(left), the feature extractor block incorporates an Edge Convolution (EdgeConv) operation\cite{DBLP:journals/corr/abs-1801-07829} followed by 3 two-dimensional convolutional (Conv2D) layers and an average pooling operation across all neighbors of each particle. The EdgeConv operation adopts a k-nearest neighbors approach with $k=20$ to extract local information for each particle based on the proximity in the $\eta-\phi$ space. All convolutional layers are implemented with stride and kernel size of 1 and are followed by a batch normalization operation and GELU activation function. Same as in Ref.~\cite{Mikuni:2021pou}, we employed two feature extractors with N=128 and N=64, respectively. 

Subsequently, we alternately stack two Particle Attention modules and two Channel Attention modules to combine both the local information of the particle representation and the global information of the jet representation. A dropout rate of 0.1 is applied to all particle attention blocks and channel attention blocks. Furthermore, inspired by Ref.~\cite{Qu:2022mxj}, we designed a channel interaction matrix based on physics principles. Then we incorporate the particle interaction matrix to the Particle Attention module and incorporate the channel interaction matrix to the Channel Attention module. For the particle interaction matrix, we utilize a 3-layer two-dimensional convolution with (32,16,8) channels with stride and kernel size of 1 to map the particle interaction matrix to a new embedding $P\times P \times N_h$, where $N_h$ is the number of heads in the particle self attention module. As for the channel interaction matrix, we utilize an upsampling operation and a 3-layer two-dimensional convolution to map the channel interaction matrix to a higher dimensional representation $N\times N$, with $N$ the input particle embedding dimension. Therefore, to process a jet of P particles, the P-DAT requires three inputs: the jet dataset, the particle interaction matrix and the jet feature interaction matrix derived from the kinetic information of each particle inside the jet. 

Next, the outputs of the Particle Attention blocks and Channel Attention blocks are concatenated, followed by an 1 dimensional Convolutional Neural Network (CNN) layer with 256 nodes and an average pooling operation across all particles. This output is then directly fed into a 3-layer MLP with (256, 128, 2) nodes, as shown in Figure.\ref{part1}(right). In addition, a batch normalization operation, a dropout rate of 0.5 and the GELU activation function are applied to the second layer. Finally, the last layer employs a softmax operation to produce the final classification scores. It is worth noting that the inclusion of class attention blocks, as described in Ref.\cite{Qu:2022mxj}, did not lead to an improvement in performance of P-DAT, as observed in our experiments.

\begin{figure}[ht]
    \centering
    \includegraphics[width=8cm,height=7cm]{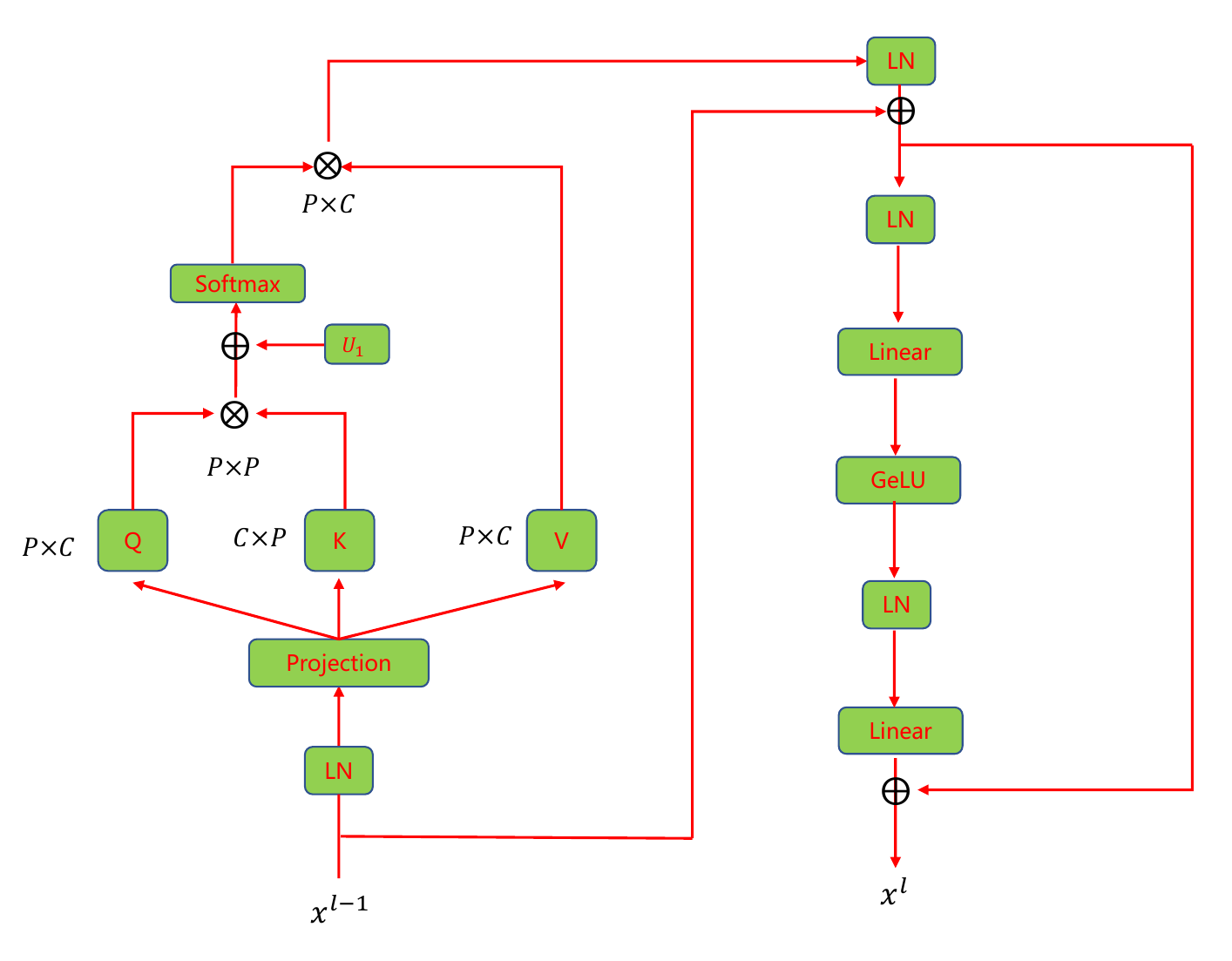}
    \caption{Illustration of the Particle Multi-head Attention Block.}
    \label{part2}
\end{figure}

\begin{figure}[ht]
    \centering
    \includegraphics[width=8cm,height=7cm]{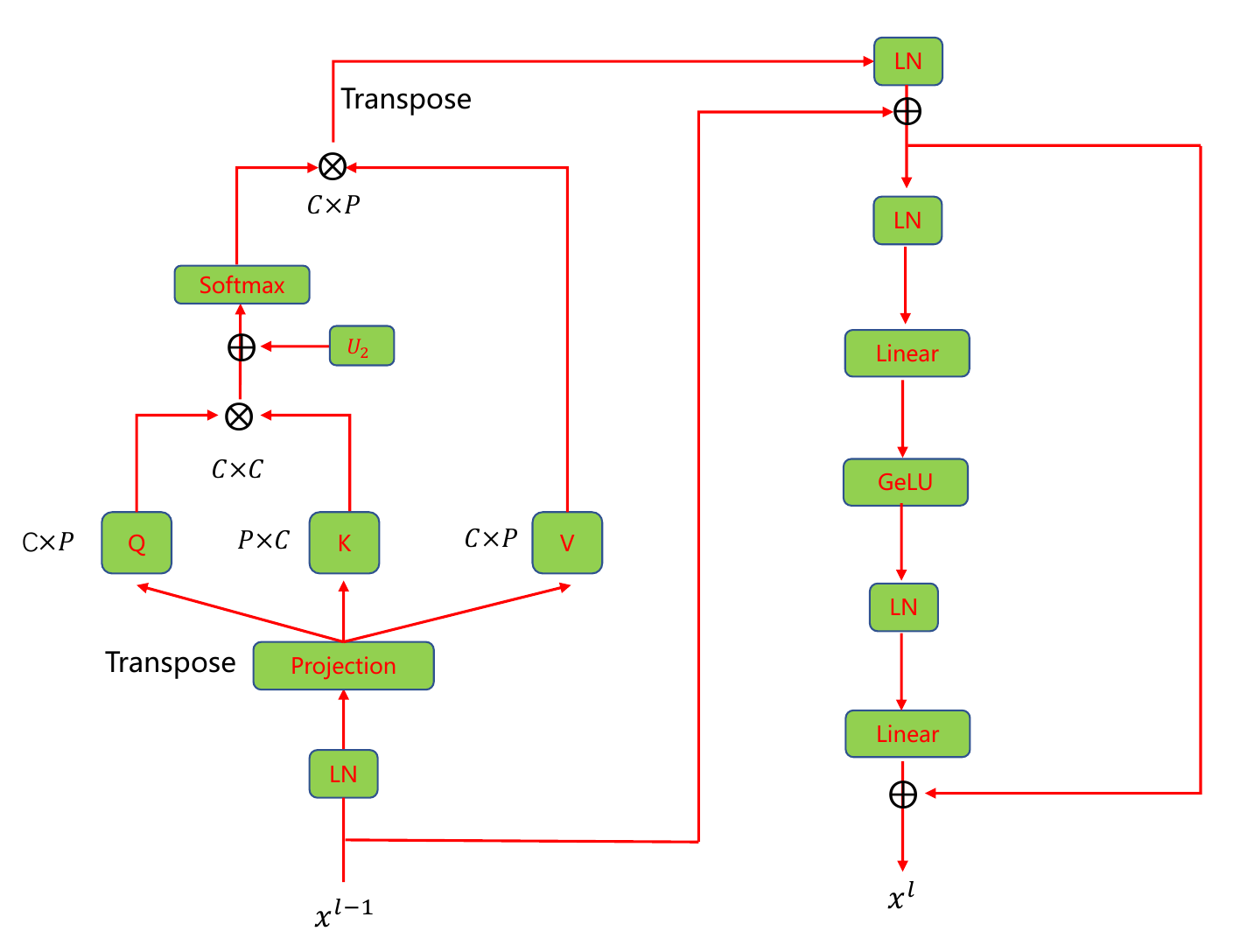}
    \caption{Illustration of the Channel Attention Block.}
    \label{part3}
\end{figure}

\subsection{Particle Attention Module}

The particle self-attention block, which is already established in the existing papers, aims to establish the relationship between all particles within the jet using an attention mechanism. As presented in Figure.\ref{part2}, three matrices, which are called query (Q), key (K), and value (V), are built from linear transformations of the original inputs. Attention weights are computed by matrix multiplication between Q and K, representing the matching between them. Same as the Particle Transformer work\cite{Qu:2022mxj}, we incorporate the particle interaction matrix $U_1$ as a bias term to enhance the scaled dot-product attention. This incorporation of particle interaction features, designed from physics principles, modifies the dot-product attention weights, thereby enhancing the expressiveness of the attention mechanism. The same $U_1$ is shared across the two particle attention blocks. After normalization, these attention weights reflect the weighted importance between each pair of particles. The self-attention is then obtained by the weighted elements of V, which results from multiplying the attention weights and the value matrix. It is important to note that $P$ represents the number of particles, and $N$ denotes the total number of features. 
The attention weights are computed as:
\begin{align}
\mathcal{A}(\mathbf{Q}, \mathbf{K}, \mathbf{V}) & = \mathrm{Concat}(\mbox{head}_1,\ldots,\mbox{head}_{N_h}) \notag \\
\text{where}~~\mbox{head}_i & = \mathrm{Attention}(\mathbf{Q}_i, \mathbf{K}_i, \mathbf{V}_i) \notag \\
& = \mathrm{softmax} \left[\frac{\mathbf{Q}_i(\mathbf{K}_i)^\mathrm{T}}{\sqrt{C_h}}+\mathbf{U_1}\right]\mathbf{V}_i
\label{eq:self-attention}
\end{align}
where $\mathbf{Q}_i=\mathbf{X}_i\mathbf{W}_i^Q$, $\mathbf{K}_i=\mathbf{X}_i\mathbf{W}_i^K$, and $\mathbf{V}_i=\mathbf{X}_i\mathbf{W}_i^V$ are $\mathbb{R}^{P \times N_h}$ dimensional visual features with $N_h$ heads, $\mathbf{X}_i$ denotes the $i_{th}$ head of the input feature and $\mathbf{W}_i$ denotes the projection weights of the $i_{th}$ head for $\mathbf{Q}, \mathbf{K}, \mathbf{V}$, and $N = C_h * N_h$. 
The particle attention block incorporates a LayerNorm layer both before and after the multi-head attention module. A two-layer MLP, with LayerNorm preceding each linear layer and GELU nonlinearity in between, follows the multi-head attention module. Residual connections are applied after the multi-head attention module and the two-layer MLP. In our study, we set $N_h=8$ and $N=64$.

\subsection{Channel Attention Module}

The main contribution of this paper is to explore the self-attention mechanism from another perspective and propose the channel-wise attention mechanism for jet tagging. Unlike the previous particle self-attention mechanism which computes the attention weights between each pair of particles, we apply attention mechanisms on the transpose of particle-level inputs and compute the attention weights between each pair of particle features. In this way, the channel-wise attention mechanism naturally capture the global interaction of each pair of particle features by taking all the particles into account, which can be viewed as the interaction of each pair of jet features. Additionally, taking inspiration from Ref.~\cite{Qu:2022mxj}, we have devised a jet feature interaction matrix based on physics principles, which can be added to enhance the expressiveness of the channel attention mechanism.

As depicted in Figure.\ref{part3}, the channel self-attention block applies attention mechanisms to the jet features, enabling interactions among the channels. To capture global information in the particle dimension, we set the number of heads to 1, where each channel represents a global jet feature. Consequently, all the channels interact with each other. This global channel attention mechanism is defined as follows:
\begin{align}
\mathcal{A}(\mathbf{Q}_i, \mathbf{K}_i, \mathbf{V}_i) & = \mathrm{softmax} \left[\frac{\mathbf{Q}_i^\mathrm{T}\mathbf{K}_i}{\sqrt{C}}+\mathbf{U_2}\right]\mathbf{V}_i^T
\label{eq:channel-attention}
\end{align}
where $\mathbf{Q}_i, \mathbf{K}_i, \mathbf{V}_i \in \mathbb{R}^{C \times P}$ are channel-wise jet-level queries, keys, and values. Note that although we perform the transpose in the channel attention block, the projection layers $\mathbf{W}$ and the scaling factor $\frac{1}{\sqrt{C}}$ are computed along the channel dimension, rather than the particle dimension. Similar as the particle self-attention block, we incorporate the designed channel interaction matrix $U_2$ as a bias term to enhance the scaled dot-product attention. The same $U_2$ matrix is shared across the two channel attention blocks. After normalization, the attention weights indicate the weighted importance of each pair of global features. The self-attention mechanism produces the weighted elements of V, obtained by multiplying the attention weights and the value matrix. Additionally, the channel attention block includes a LayerNorm layer before and after the attention module, followed by a two-layer MLP. Each linear layer is preceded by a LayerNorm layer, and a GELU nonlinearity is applied between them. Residual connections are added after the channel attention module and the two-layer MLP.

\subsection{Combination of Particle Attention module and Channel Attention module}

Throughout the whole architecture, all the Particle Attention modules and the Channel Attention modules are stacked while maintaining a consistent feature dimension of $N=64$. The Channel Attention module captures global information and interactions, while the Particle Attention module extracts local information and interactions. In the context of the channel self-attention mechanism, a $C\times C$-dimensional attention map is computed, involving all the particles, resulting in a computation of the form $(C\times P) \cdot (P\times C)$. This global attention map enables the channel attention module to dynamically fuse multiple global perspectives of the jet. Subsequently, a transpose operation is performed, yielding outputs with new channel information, which are then passed to the subsequent Particle Attention module. Conversely, in the particle self-attention mechanism, a $P\times P$-dimensional attention map is computed by considering all particle features, resulting in a computation of the form $(P\times C) \cdot (C\times P)$. This local attention map empowers the particle attention module to dynamically fuse multiple local views of the jet, generating new particle features and passing the information to the following Channel Attention module. By alternatively applying these two types of modules, the local information and global information can complement each other.



\subsection{Model Implementation}

The PYTORCH\cite{NIPS2019_9015} deep learning framework is utilized to implement the model architecture with the CUDA platform. The training and evaluation steps are accelerated using a NVIDIA GeForce RTX 3070 GPU for acceleration. We adopt the binary cross-entropy as the loss function. To optimize the model parameters, we employ the AdamW optimizer\cite{loshchilov2019decoupled} with an initial learning rate of 0.0005, which is determined based on the gradients calculated on a mini-batch of 64 training examples. The network is trained up to 100 epochs, with the learning rate decreasing by a factor of 2 every 10 epochs to a minimal of $10^{-6}$. In addition, we employ the early-stopping technique to prevent over-fitting. 

Furthermore, as mentioned in Ref.~\cite{Qu:2022mxj}, the introduction of the pairwise interaction matrix based on physics principle significantly increases the computational time and memory consumption, therefore limiting the number of pairwise interaction matrix which is the prior knowledge based on physics principle. In this paper, in order to address the memory issue caused by huge input data, we implemented the Chunk Loading strategy, a commonly used technique in the field of deep learning for data loading. This approach entails continuously importing and deleting data during the training, validation and test process, enabling us to train our model on a large dataset while mitigating the memory load. We give a detailed description of this approach in the following:

Within a loop, input data batches are dynamically loaded for training, validation, and test. Each batch contains 1280 events. Regardless of whether it's for training, validation, or testing, the data loading process remains consistent. This uniformity ensures that the iteration counts for training, validation, and testing may vary, but the data-handling approach remains the same. During each iteration, we employ NumPy's memory-mapped file access to efficiently retrieve training data, corresponding labels, particle interaction matrices, and jet interaction matrices. Once this batch is processed for training/validation/testing, the loaded data is subsequently removed to free up memory resources. Subsequently, we proceed to load the next batch of data for next iteration. This method significantly reduces memory consumption by allowing us to access the necessary data without the need to load the entire dataset into memory all at once. This strategic approach not only optimizes memory utilization but also effectively mitigates the challenges associated with handling substantial input data. It allows us to train our model efficiently while preventing memory exhaustion.

\section{Results of Jet Classification}
\label{sec:Classification}

The P-DAT architecture is designed to process input data consisting of particles inside the jets. Based on the point cloud representation, we regard each constituent particle as a point in the $\eta-\phi$ space and the whole jet as a point cloud. To ensure consistency and facilitate meaningful comparisons, we first sorted the particles inside the jets by transverse momentum and a maximum of 100 particles per jet are employed. The input jet is truncated if the particle number inside the jet is more than 100 and the input jet is zero-padded up to the 100 if fewer than 100 particles are present. In this process, the zero-padded constituent particles were directly introduced as zeros into the model, without the utilization of any additional masking. This selection of 100 particles is sufficient to cover the vast majority of jets contained within all datasets, ensuring comprehensive coverage. Each jet is characterized by the 4-momentum of its constituent particles. Based on this information, we reconstructed 7 features for each particle. Additionally, for the quark-gluon dataset, we included the Particle Identification (PID) information as the 8-th feature. These features are as follows:

\begin{eqnarray}
\left\{
\begin{aligned}
\text{log $E$} ,\ \text{log $p_\text{T}$} ,\ \frac{p_{\text{T}}}{p_{\text{TJ}}} ,\ \frac{E}{E_J} ,\ \Delta \eta \  \Delta \phi ,\ \Delta R , \ \text{PID}
\end{aligned}
\right\}.
\end{eqnarray}

For the pairwise particle interaction matrix, we adopt the same four features as employed in Refs. \cite{Qu:2022mxj, Dreyer:2020brq}. Additionally, we include the difference in transverse momentum as an additional feature. To summarize, we calculated the following 5 features for any pair of particles a and b with four-momentum $p_a$ and $p_b$, respectively:

\begin{align}
\begin{split}
    \Delta R&= \sqrt{(y_a - y_b)^2 + (\phi_a - \phi_b)^2}, \\
    k_{\text{T}} &= \min(p_{\text{T},a}, p_{\text{T},b}) \Delta, \\
    z &= \min(p_{\text{T},a}, p_{\text{T},b}) / (p_{\text{T},a} + p_{\text{T},b}), \\
    m^2 &= (E_a+E_b)^2 - \|\mathbf{p}_{a}+\mathbf{p}_{b}\|^2, \\
    \Delta p_{\text{T}}&= |p_{\text{T},a}-p_{\text{T},b}| 
\label{eq:interaction}
\end{split}
\end{align}

where $y_i$ represents the rapidity, $\phi_i$ denotes the azimuthal angle, $p_{\text{T},i} = (p_{x, i}^2+p_{y, i}^2)^{1/2}$ denotes the transverse momentum, $\mathbf{p}_i=(p_{x,i}, p_{y,i}, p_{z,i})$ represents the momentum 3-vector and $\|\cdot\|$ is the norm, for $i=a$, $b$. As mentioned in Ref.\cite{Qu:2022mxj}, we take the logarithm and use $(\ln \Delta, \ln k_{\text{T}}, \ln z, \ln m^2, \ln \Delta p_{\text{T}})$ as the interaction features for each particle pair to avoid the long tail problem. Moreover, apart from the 5 interaction features, we design one more feature for the Quark-Gluon benchmark dataset, defined as $\delta_{i,j}$, where i and j are the Particle Identification of the particles a and b.

Furthermore, as mentioned in Section.~\ref{sec:Architecture}, we have designed a pairwise jet feature interaction matrix, drawing inspiration from the work Ref.~\cite{Qu:2022mxj}. The list of all jet features used in this study is presented below. Note that all the jet features are calculated based on the four-momentum of all the constituent particles within the jet. The interaction matrix is constructed based on a straightforward yet effective ratio relationship, as illustrated in Table.\ref{U2}.

\begin{eqnarray}
\Bigg\{
\begin{aligned}
&\text{E} ,\  \text{$p_\text{T}$} ,\ \sum p_{Tf} ,\ \sum E_f ,
&\overline{\Delta \eta}, \  \overline{\Delta \phi} ,\ \overline{\Delta R} , \ \text{PID}
\end{aligned}
\Bigg\}.
\end{eqnarray}

\begin{table*}[ht]
\setlength{\tabcolsep}{3pt} 
\renewcommand{\arraystretch}{1.8} 
\begin{center}
\begin{tabular}{|c|c|c|c|c|c|c|c|c|c|c|c|}
\cline{1-9} I  & E & $p_T$ & $\sum p_{Tf}$  & $\sum E_f$ & $\overline{\Delta \eta}$  & $\overline{\Delta \phi}$ & $\overline{\Delta R}$  & PID \\
\cline{1-9} E  & 1 & $\frac{p_{T}}{E}$ & 0 & 1 & 0 & 0 & 0 & $\frac{E_{PID}}{E}$\\
\cline{1-9} $p_T$ & $\frac{p_{T}}{E}$ & 1 & 1  & 0 & 0 & 0  & 0 &  $\frac{p_{TPID}}{p_T}$  \\
\cline{1-9}  $\sum p_{Tf}$ & 0 & 1 & 1 & 0 & 0 & 0 & 0 & $p_{TfPID}$ \\
\cline{1-9}  $\sum E_f$ & 1 & 0 & 0 & 1 & 0 & 0 & 0 & $E_{fPID}$ \\
\cline{1-9}  $\overline{\Delta \eta}$ & 0 & 0 & 0 & 0 & 1 & 0 & $\frac{\overline{\Delta \eta}}{\overline{\Delta R}}$ & $\overline{\Delta \eta}_{PID}$ \\
\cline{1-9}  $\overline{\Delta \phi}$ & 0 & 0 & 0 & 0 & 0 & 1 & $\frac{\overline{\Delta \phi}}{\overline{\Delta R}}$ & $\overline{\Delta \phi}_{PID}$ \\
\cline{1-9}  $\overline{\Delta R}$ & 0 & 0 & 0 & 0 & $\frac{\overline{\Delta \eta}}{\overline{\Delta R}}$ & $\frac{\overline{\Delta \phi}}{\overline{\Delta R}}$ & 1 & $\overline{\Delta R}_{PID}$ \\
\cline{1-9}  PID & $\frac{E_{PID}}{E}$ & $\frac{p_{TPID}}{p_T}$ & $p_{TfPID}$ & $E_{fPID}$ & $\overline{\Delta \eta}_{PID}$ & $\overline{\Delta \phi}_{PID}$ & $\overline{\Delta R}_{PID}$  & 1 \\
\cline{1-9} \end{tabular}
\caption{The jet feature pairwise interaction matrix used as the inputs for the P-DAT. Here PID represents the Particle Identification.}
\label{U2}
\end{center}
\end{table*}

To provide a clearer explanation of the concept of the jet feature pairwise interaction matrix, we will now present a detailed description. The first variable $E$ represents the energy of the input jet. $p_{\text{T}}$ denotes the transverse momentum of the input jet, while $\sum p_{Tf}$ and $\sum E_f$ represent the sum of the transverse momentum fractions and the energy fractions of all the constituent particles inside the input jet, respectively. Additionally, $\overline{\Delta \eta}$, $\overline{\Delta \phi}$ and $\overline{\Delta R}$ correspond to the transverse momentum weighted sum of the $\Delta \eta$, $\Delta \phi$, $\Delta R$ of all the constituent particles inside the input jet, respectively. Here $\Delta \eta$, $\Delta \phi$ and $\Delta R$ refer to the distances in the $\eta-\phi$ space between each constituent particle and the input jet. Furthermore, for the quark-gluon dataset, we incorporated the 8th feature based on the Particle Identification information. It represents the particle identification associated with the specific particle type whose sum of transverse momentum accounts for the largest proportion of the entire jet transverse momentum. The entire jet feature pairwise interaction matrix is defined as a symmetric block matrix with diagonal ones. For convenience, we named $\{\text{E} ,\ 
\text{$p_\text{T}$} ,\ \sum p_{Tf} ,\ \sum E_f\}$ as variable set 1 and $\{\overline{\Delta \eta}, \  \overline{\Delta \phi} ,\ \overline{\Delta R}\}$ as variable set 2. We build the pairwise interactions among variable set 1 and variable set 2, respectively. Firstly, we employ a ratio relationship to define the interaction between E and $\text{$p_\text{T}$}\}$. Additionally, we establish that the interaction between $\sum E_f$ and E is 1, while no interactions exist between $\sum E_f$ and any other variables, except for E and Particle Identification. Similarly, we define the interaction between $\sum p_{Tf}$ and $p_T$ as 1, with no interactions between $\sum p_{Tf}$ and any other variables, except for $p_T$ and Particle Identification.

Secondly, we apply a ratio relationship to define the interaction between $\overline{\Delta R}$ and $\{\overline{\Delta \eta}, \overline{\Delta \phi} \}$, while no interaction is specified between $\overline{\Delta \eta}$ and $\overline{\Delta \phi}$. Finally, we determine the interactions between Particle Identification and all other variables as the ratio of the sum of the corresponding variables of the particles associated with the Particle Identification to the variable of the jet.

\subsection{Quark/Gluon Discrimination}
\label{sec:qg}

The Quark-Gluon benchmark dataset\cite{Komiske:2018cqr} was produced using Pythia8\cite{Sjostrand:2014zea} without detector simulation. It includes quark-initiated samples $q\overline{q} \rightarrow{Z\rightarrow{\nu\overline{\nu}}+(u,d,s)}$ as signal and gluon-initiated data $q\overline{q} \rightarrow{Z\rightarrow{\nu\overline{\nu}}+g}$ as background. 
Jet clustering was performed using the anti-kT algorithm with R = 0.4. Only jets with transverse momentum $p_T \in$ [500, 550] GeV and rapidity $|y| < 1.7$ were selected for further analysis. Each particle within the dataset comprises not only the four-momentum, but also the particle identification information, which classifies the particle type as electron, muon, charged hadron, neutral hadron, or photon. The official dataset compromises of 1.6M training events, 200k validation events and 200k test events, respectively.  In this paper, we focused on the leading 100 constituents within each jet, utilizing their four-momenta and particle identification information for training purposes. For jets with fewer than 100 constituents, zero-padding was applied. For each particle, a set of 8 input features was used, based solely on the four-momenta and identification information of the particles clustered within the jet. The accuracy, area under the curve (AUC), and background rejection results are presented in Table. \ref{tab:results_qg}.

\begin{table}[htb]
    \centering
    \caption{Comparison between the performance reported  for P-DAT and existing classification algorithms on the quark-gluon discrimination dataset. The uncertainty is calculated by taking the standard deviation of 5 training runs with different random weight initialization.}
    \label{tab:results_qg}
    \setlength{\tabcolsep}{3pt}
	\begin{tabular}{lccccc}
             \hline
          &  Accuracy &AUC & Rej$_{50\%}$  & Rej$_{30\%}$ \\
            \hline
            ResNeXt-50 \cite{Qu:2019gqs} & 0.821 & 0.9060 & 30.9 & 80.8 \\
            P-CNN \cite{Qu:2019gqs} & 0.827 & 0.9002 & 34.7 & 91.0 \\
            PFN \cite{Komiske:2018cqr} & - & 0.9005 & 34.7$\pm$0.4 & - \\
            ParticleNet-Lite \cite{Qu:2019gqs} & 0.835 & 0.9079 &  37.1 & 94.5 \\
            ParticleNet \cite{Qu:2019gqs} & 0.840 & 0.9116 & 39.8$\pm$0.2 & 98.6$\pm$1.3\\
            ABCNet \cite{Mikuni:2020wpr}& 0.840 & 0.9126 & 42.6$\pm$0.4 & 118.4$\pm$1.5 \\
            SPCT \cite{Mikuni:2021pou} & 0.815 & 0.8910 & 31.6$\pm$0.3 & 93.0$\pm$1.2 \\
            PCT \cite{Mikuni:2021pou} & 0.841 & 0.9140 & 43.2$\pm$0.7 & 118.0$\pm$2.2 \\
            LorentzNet \cite{Gong:2022lye} & 0.844 & 0.9156 & 42.4$\pm$0.4 & 110.2$\pm$1.3 \\
            ParT \cite{Qu:2022mxj} & 0.849 & 0.9203 & 47.9$\pm$0.5 & 129.5$\pm$0.9 \\
            \hline
            P-DAT & 0.839 & 0.9092 & 39.2$\pm$0.6 & 95.1$\pm$1.3 \\
            \hline
	\end{tabular}
\end{table}

From Table. \ref{tab:results_qg}, we can see that in the context of the Quark/Gluon Discrimination task, P-DAT exhibits powerful classification performance, surpassing the majority of models while falling slightly behind other two transformer-based models, PCT and ParT. The superior results of ParT can be attributed to its significantly more complex architecture with a total of L = 8 particle attention blocks and 2 class attention blocks. The model complexity of ParT exceeds the P-DAT model by a substantial margin. As for the PCT model, all self-attention layers employ query, key, and value matrices obtained through one-dimensional convolutional layers, resulting in a larger number of FLOPs compared to our model. P-DAT strikes a favorable balance between performance and model complexity. Additionally, our P-DAT model incorporates the Channel Attention module, offering greater flexibility in leveraging abundant jet information compared to the other two methods.

\subsection{Top Tagging}
\label{sec:top}

The benchmark dataset\cite{Benato:2021olt} used for top tagging comprises hadronic tops as the signal and QCD di-jets as the background. Pythia8\cite{Sjostrand:2014zea} was employed for event generation, while Delphes\cite{deFavereau:2013fsa} was utilized for detector simulation. All the particle-flow constituents were clustered into jets using the anti-kT algorithm\cite{Cacciari:2008gp} with a radius parameter of R = 0.8. Only jets with transverse momentum $p_T \in$ [550, 650] GeV and rapidity $|y| < 2$ were included in the analysis. The official dataset contains 1.2M training events, 400k validation events and 400k test events, respectively. Only the energy-momentum 4-vectors for each particles inside the jets are provided. In this paper, the leading 100 constituent four-momenta of each jet were utilized for training purposes. For jets with fewer than 100 constituents, zero-padding was applied. For each particle, a set of 7 input features based solely on the four-momenta of the particles clustered inside the jet was utilized. The accuracy, area under the curve (AUC), and background rejection results can be found in Table. \ref{tab:results_top}.

\begin{table}[htb]
    \centering
    \caption{Comparison between the performance reported  for P-DAT and existing classification algorithms on the top tagging dataset. The uncertainty is calculated by taking the standard deviation of 5 training runs with different random weight initialization.}
    \label{tab:results_top}
    \setlength{\tabcolsep}{3pt}
	\begin{tabular}{lcccc}
            \hline 
          &  Accuracy &AUC & Rej$_{50\%}$  & Rej$_{30\%}$ \\
            \hline
            ResNeXt-50 \cite{Qu:2019gqs} & 0.936 & 0.9837 & 302$\pm$5 & 1147$\pm$58 \\
            P-CNN \cite{Qu:2019gqs} & 0.930 & 0.9803 & 201$\pm$4 & 759$\pm$24 \\
            PFN \cite{Komiske:2018cqr} & - & 0.9819 & 247$\pm$3 & 888$\pm$17 \\
            ParticleNet-Lite \cite{Qu:2019gqs} & 0.937 & 0.9844 &  325$\pm$5 & 1262$\pm$49 \\
            ParticleNet \cite{Qu:2019gqs} & 0.940 & 0.9858 & 397$\pm$7 & 1615$\pm$93\\
            JEDI-net  \cite{Moreno:2019bmu}  & 0.9263 & 0.9786 & - & 590.4 \\
            SPCT \cite{Mikuni:2021pou} & 0.928 & 0.9799 & 201$\pm$9 & 725$\pm$54 \\
            PCT \cite{Mikuni:2021pou} & 0.940 & 0.9855 & 392$\pm$7 & 1533$\pm$101 \\
            LorentzNet \cite{Gong:2022lye} & 0.942 & 0.9868 & 498$\pm$18 & 2195$\pm$173 \\
            ParT \cite{Qu:2022mxj} & 0.940 & 0.9858 & 413$\pm$16 & 1602$\pm$81 \\
            \hline
            P-DAT & 0.932 & 0.9768 & 228$\pm$8 & 876$\pm$39 \\
            \hline
	\end{tabular}
\end{table}

From Table. \ref{tab:results_top}, a similar pattern emerges when analyzing the performance of models in the top tagging task. P-DAT exhibits competitive classification performance. While other two transformer-based models, PCT and ParT, achieve modestly enhanced performance, especially in terms of background rejection rates, which reach nearly twice that of our P-DAT model, this advantage comes at the cost of increased model complexity and resource demands.

Furthermore, given that our P-DAT model includes the Channel Attention module and considering the distinct jet substructure characteristics observed in boosted top jets and boosted QCD jets, we have the opportunity to formulate a set of jet substructure variables and develop an additional self-attention module to calculate attention weights for every pair of these jet substructure variables. The resulting attention weight matrix can be employed as a bias term to augment channel scaled dot-product attention. This can be an interesting research direction in the future to enhance the performance of top tagging. While we acknowledge that ParticleNet Lite achieves higher background rejection rates with smaller model complexity regarding Top Tagging task, we believe that the adaptability and innovation inherent in the P-DAT model, combining the global jet information and local particle information, pave the way for exciting possibilities in this field.

\section{Computational Complexity}
\label{sec:Complexity}

In addition to evaluating the algorithm's performance, it's crucial to consider the computational cost involved. To gauge the computational resources needed for assessing each model, we calculate both the number of trainable parameters and the number of floating-point operations (FLOPs). Table.~\ref{tab:results_cost} presents a comparative analysis of these factors across various algorithms.

\begin{table}[htb]
    \centering
    \caption{Comparison between the number of trainable weights and floating point operations (FLOPs) reported for P-DAT and existing classification algorithms.}
    \label{tab:results_cost}
    \setlength{\tabcolsep}{3pt}
	\begin{tabular}{lccccc}
             \hline 
          & Parameters & FLOPs  \\
            \hline
            ResNeXt-50 \cite{Qu:2019gqs} & 1.46M & -  \\
            P-CNN \cite{Qu:2019gqs} & 354k & 15.5M \\
            PFN \cite{Komiske:2018cqr} & 86.1k & 4.62M \\
            ParticleNet-Lite \cite{Qu:2019gqs} & 26k & - \\
            ParticleNet \cite{Qu:2019gqs} & 370k & 540M \\
            ABCNet \cite{Mikuni:2020wpr}& 230k & - \\
            SPCT \cite{Mikuni:2021pou} & 7k & 2.4M \\
            PCT \cite{Mikuni:2021pou} & 193.3k & 266M \\
            LorentzNet \cite{Gong:2022lye} & 224k & - \\
            ParT \cite{Qu:2022mxj} & 2.13M & 260M \\
            \hline
            P-DAT & 498k & 144M \\
            \hline
	\end{tabular}
\end{table}

In the context of computational complexity comparison among various models, our P-DAT model emerges as a notable candidate. While the number of P-DAT trainable parameters is increased by more than 2.6 times compared to PCT, the number of floating point operations (FLOPs) is actually 45$\%$ lower. Notably, when compared to ParticleNet, PCT, and ParT, P-DAT features the smallest FLOPs. P-DAT distinguishes itself by maintaining a comparatively modest parameter count at 498k while offering a reasonable level of computational efficiency with 144M FLOPs. This balance between model complexity and computational demands positions P-DAT as an attractive choice for practical applications, where it can potentially deliver competitive performance with fewer computational resources, making it a promising option for deployment and further research.

\section{Conclusion}
\label{sec:Summary}

In this study, we introduced the Particle Dual Attention Transformer (P-DAT) as an innovative model architecture for jet tagging. We designed the Channel Attention module and alternately employed the Particle Attention module and the Channel Attention module to capture both jet-level global information and particle-level local information, while maintaining computational efficiency. Additionally, we incorporate both the pairwise particle interactions and the pairwise jet feature interactions in the attention mechanism. We evaluate the P-DAT architecture on the classic top tagging task and the quark-gluon discrimination task and achieve competitive results compared to other benchmark strategies. Notably, our P-DAT maintains a relatively modest parameter count 498k while simultaneously delivering a reasonable level of computational efficiency with 144M FLOPs, which strikes a balance between computational complexity and model performance.
Besides, given the substantial computational demands posed by introducing a pairwise interaction matrix based on physics principles, which can impact both time and memory resources, we have introduced the Chunk Loading strategy which involves dynamic data import and deletion throughout the training, validation, and testing phases, effectively addressing memory usage constraints.

Finally, Channel Attention module opens up more possibilities for future exploration. For instance, in this study we proposed the Channel Attention module and designed the jet feature interaction matrix as our primary contributions. As an alternative approach to utilizing simple ratio-based interaction matrix, we could explore the possibility of constructing a dedicated attention module for jet features. By incorporating the resulting attention weight matrix into the Channel Attention module, we may potentially enhance performance. This strategy offers the advantage of incorporating valuable supplementary jet information and leveraging the intrinsic patterns within jet features revealed by the jet feature attention mechanism.

\begin{acknowledgements}
This work of Daohan Wang is funded by the National Research Foundation of Korea, Grant No. NRF-2022R1A2C1007583. The work of Minxuan He is supported by the Fundamental Research Funds for the Central Universities.
\end{acknowledgements}

\bibliography{ref}

\end{document}